\documentclass[aps,pra,reprint,showpacs,superscriptaddress,nofootinbib,twocolumn]{revtex4-2}
\usepackage[dvips]{graphicx} 
\usepackage{amsfonts,amscd,amsmath,amsthm}
\usepackage{enumerate}
\usepackage{epsfig}
\usepackage{subfigure}
\usepackage{xcolor}
\usepackage[colorlinks = true]{hyperref}
\usepackage{physics}
\usepackage{booktabs}
\usepackage{epstopdf}
\usepackage{framed}
\usepackage{multirow}
\usepackage{color}
\usepackage{longtable}
\usepackage{comment}
\usepackage[ruled,vlined]{algorithm2e}
\usepackage[most]{tcolorbox}
\usepackage{float}

\graphicspath{{./figure/}}

\usepackage{tikz}
\usetikzlibrary{tikzmark, calc, fit, positioning}
\usetikzlibrary{shapes,arrows}
\usetikzlibrary{quantikz}

\newcommand{\mysection}[1]{\textbf{#1}.}
\newtcolorbox[auto counter]{mybox}[2][]{
	enhanced,
	breakable,
	colback=blue!5!white,
	colframe=blue!75!black,
	fonttitle=\bfseries,
	title=Box \thetcbcounter: #2,#1
}

\begin{document}
\onecolumngrid
\title{Interplay of Quantum Resources in Nonlocality Tests}

\author{Hai-Hao Dong}
\affiliation{Hefei National Research Center for Physical Sciences at the Microscale and School of Physical Sciences, University of Science and Technology of China, Hefei, China.}
\affiliation{Shanghai Research Center for Quantum Science and CAS Center for Excellence in Quantum Information and Quantum Physics, University of Science and Technology of China, Shanghai, China.}
\affiliation{Hefei National Laboratory, University of Science and Technology of China, Hefei, China.}
\author{Yuwei Zhu}
\affiliation{Center for Quantum Information, Institute for Interdisciplinary Information Sciences, Tsinghua University, Beijing 100084, P. R. China}
\affiliation{Yau Mathematical Sciences Center, Tsinghua University, Beijing 100084, P. R. China}
\author{Su-Yi Cheng}
\affiliation{Hefei National Research Center for Physical Sciences at the Microscale and School of Physical Sciences, University of Science and Technology of China, Hefei, China.}
\affiliation{Shanghai Research Center for Quantum Science and CAS Center for Excellence in Quantum Information and Quantum Physics, University of Science and Technology of China, Shanghai, China.}
\affiliation{Hefei National Laboratory, University of Science and Technology of China, Hefei, China.}
\author{Xingjian Zhang}
\affiliation{Hefei National Research Center for Physical Sciences at the Microscale and School of Physical Sciences, University of Science and Technology of China, Hefei, China.}
\affiliation{Shanghai Research Center for Quantum Science and CAS Center for Excellence in Quantum Information and Quantum Physics, University of Science and Technology of China, Shanghai, China.}
\affiliation{Center for Quantum Information, Institute for Interdisciplinary Information Sciences, Tsinghua University, Beijing 100084, P. R. China}
\author{Cheng-Long Li}
\affiliation{Hefei National Research Center for Physical Sciences at the Microscale and School of Physical Sciences, University of Science and Technology of China, Hefei, China.}
\affiliation{Shanghai Research Center for Quantum Science and CAS Center for Excellence in Quantum Information and Quantum Physics, University of Science and Technology of China, Shanghai, China.}
\affiliation{Hefei National Laboratory, University of Science and Technology of China, Hefei, China.}
\author{Ying-Zhao Li}
\affiliation{Hefei National Research Center for Physical Sciences at the Microscale and School of Physical Sciences, University of Science and Technology of China, Hefei, China.}
\affiliation{Shanghai Research Center for Quantum Science and CAS Center for Excellence in Quantum Information and Quantum Physics, University of Science and Technology of China, Shanghai, China.}
\affiliation{Hefei National Laboratory, University of Science and Technology of China, Hefei, China.}

\author{Hao Li}
\affiliation{Shanghai Key Laboratory of Superconductor Integrated Circuit Technology, Shanghai Institute of Microsystem and Information Technology, Chinese Academy of Sciences, Shanghai 200050, China}

\author{Lixing You}
\affiliation{Shanghai Key Laboratory of Superconductor Integrated Circuit Technology, Shanghai Institute of Microsystem and Information Technology, Chinese Academy of Sciences, Shanghai 200050, China}

\author{Xiongfeng Ma}
\email{xma@tsinghua.edu.cn}
\affiliation{Center for Quantum Information, Institute for Interdisciplinary Information Sciences, Tsinghua University, Beijing 100084, P. R. China}

\author{Qiang Zhang}
\email{qiangzh@ustc.edu.cn}
\affiliation{Hefei National Research Center for Physical Sciences at the Microscale and School of Physical Sciences, University of Science and Technology of China, Hefei, China.}
\affiliation{Shanghai Research Center for Quantum Science and CAS Center for Excellence in Quantum Information and Quantum Physics, University of Science and Technology of China, Shanghai, China.}
\affiliation{Hefei National Laboratory, University of Science and Technology of China, Hefei, China.}

\author{Jian-Wei Pan}
\email{pan@ustc.edu.cn}
\affiliation{Hefei National Research Center for Physical Sciences at the Microscale and School of Physical Sciences, University of Science and Technology of China, Hefei, China.}
\affiliation{Shanghai Research Center for Quantum Science and CAS Center for Excellence in Quantum Information and Quantum Physics, University of Science and Technology of China, Shanghai, China.}
\affiliation{Hefei National Laboratory, University of Science and Technology of China, Hefei, China.}
\begin{abstract}

Nonlocality, evidenced by the violation of Bell inequalities, not only signifies entanglement but also highlights measurement incompatibility in quantum systems. Utilizing the generalized Clauser-Horne-Shimony-Holt (CHSH) Bell inequality, our high-efficiency optical setup achieves a loophole-free violation of $2.0132$. This result provides a device-independent lower bound on entanglement, quantified as the entanglement of formation at $0.0159$. Moreover, by tuning the parameters of the generalized Bell inequality, we enhance the estimation of measurement incompatibility, which is quantified by an effective overlap of $4.3883 \times 10^{-5}$. To explore the intricate interplay among nonlocality, entanglement, and measurement incompatibility, we generate mixed states, allowing for flexible modulation of entanglement via fast switching among the four Bell states using Pockels cells, achieving a fidelity above $99.10\%$. Intriguingly, our results reveal a counterintuitive relationship where increasing incompatibility initially boosts nonlocality but eventually leads to its reduction. Typically, maximal nonlocality does not coincide with maximal incompatibility. This experimental study sheds light on the optimal management of quantum resources for Bell-inequality-based quantum information processing.

\end{abstract}

\maketitle
\onecolumngrid
\mysection{Introduction}
In response to Einstein, Podolsky, and Rosen's argument that quantum physics is an ``incomplete" theory based on the local realism assumption~\cite{Einstein1935}, 
Bell proved that the predictions of quantum theory
are incompatible with those of any physical theory satisfying local hidden-variable (LHV) models. This incompatible difference involves three quantum resources: \textit{Bell nonlocality}, \textit{entanglement}, and \textit{measurement incompatibility}~\cite{RevModPhys.QRT}. Specifically, we speak of Bell nonlocality when LHV models do not hold~\cite{scarani2019bell}, which is indicated by the violation of Bell inequalities~\cite{brunner2014bell}. Entanglement characterizes the nonclassical nature of a composite system state that cannot be decomposed into a product of individual subsystem states. 
Incompatibility goes back to Heisenberg's uncertainty principle~\cite{heisenberg1927anschaulichen}, indicating the presence of quantum measurements that cannot be simultaneously implemented. While both entanglement and measurement incompatibility are necessary conditions for nonlocality, the significance of measurement incompatibility is often neglected, leaving a research gap in this domain. In addition to their significance for testing quantum foundations, these three quantum resources have broad applications in quantum cryptography, including quantum key distribution~\cite{PhysRevLett1991Ekert, mayers1998quantum,Barrett2005No,Acin2007device,Liu2022Toward,nadlinger2022experimental,zhang2022device}, quantum random number generation~\cite{pironio2010random,Colbeck2011Private,bierhorst2018experimentally,liu2018device,Zhang2020Experimental,Li2021Experimental,shalm2021device,li2023device}.

Given the essential role that a sufficient level of nonlocality, entanglement, and incompatibility plays in quantum protocols, quantifying these resources is fundamental and significant. Moreover, finding the quantitative relation among them is also important but remains unclear. Thanks to the necessity between entanglement, incompatibility and nonlocality, nonlocality can provide valid estimations of these two resources, even in the presence of untrusted devices, known as \textit{device-independent} (DI) quantification~\cite{Bennett1996Mixed,Vedral1997Quantifying,mayers1998quantum,Acin2007device,Verstraete2002Entanglement,Liang2011Semi,Moroder2013Device,Toth2015Evaluating,Arnonfriedman2018Noise,Chen2018Exploring,Arnon-Friedman_2019Device}. Back to the relation among the three, we intuitively infer that entanglement (incompatibility) exhibits a monotonic relationship with nonlocality while holding the other quantity constant. However, the relationship among diverse quantum properties can be counterintuitive~\cite{Acin2012Randomness,Wooltorton2024Device}. Theoretical research has demonstrated that the connection between state entanglement and measurement incompatibility for a given nonlocal behavior is not a straightforward trade-off~\cite{zhu2023entanglement}.
This indicates that different couples of entanglement and incompatibility can lead to equivalent nonlocality, enabling the optimal allocation of these quantum resources in Bell-inequality-based quantum information processing~\cite{tendick2022quantifying}.

In experiments, to observe the quantitative relationship, Bell violations in different combinations of entanglement and measurement incompatibility are needed. To make the results more general, these violations should be extended to generalized Bell inequalities. Achieving these outcomes convincingly, in a loophole-free manner~\cite{hensen2015loophole,Shalm2015Strong,Giustina2015Significant,Li2018Test}, necessitates a spacelike separated system characterized by high detection efficiency and state fidelity. Additionally, a method for preparing mixed states with adjustable levels of entanglement is also required.

In this work, the experiments are firstly conducted in a device-independent manner by closing the locality and detection loopholes with the state-of-art efficiency in a spacelike separated Bell test setup.
We considered two specific entanglement measures, including the entanglement of formation (EOF)~\cite{Bennett1996Mixed} and the negativity of entanglement~\cite{Vidal2002Computable}, as they can be quantitive estimated by a given Bell inequality violation. We also estimate the tight bound of measurement incompatibility defined by effective overlap~\cite{Tomamichel2013The}. 

To further investigate the interplay among nonlocality, entanglement, and incompatibility,  we use a mixed-state preparation technique by fast switching between Bell bases with high frequency, efficiency, and fidelity. This approach allows the entanglement of the Bell diagonal states to be precisely controlled by electronic parameters.
\\
\\
\mysection{Resources Quantification via Nonlocality Test}
For systematical investigation, we use a set of generalized CHSH-type Bell inequalities with an extra parameter $\alpha$~\cite{Acin2012Randomness}, 
\begin{equation}
\label{eq:CHSH_type}
   S_{\alpha}=\alpha\langle\hat{A_0}\otimes \hat{B_0}\rangle+\alpha \langle\hat{A}_0\otimes \hat{B}_1\rangle+\langle\hat{A}_1\otimes \hat{B}_0\rangle-\langle\hat{A}_1\otimes \hat{B}_1\rangle,
\end{equation}
where $\langle\hat{A}_{x}\otimes\hat{B}_{y}\rangle=\sum_{a,b}ab\cdot p(a,b|x,y)$, $a,b\in\{\pm 1\}$. Here $p(a,b|x,y)$ denotes the outcome probability conditioned on the measurements $\hat{A}_{x}$ and $\hat{B}_{y}$ by Alice and Bob according to their measurement inputs, $x,y\in\{0,1\}$, respectively. The parameter $\alpha\geq 1$ is an extra term partially tuning the contribution of inputs. When $\alpha=1$, Eq.~\eqref{eq:alpha_CHSH} reduces to the original CHSH value obtained in the game.

Firstly, we use EOF as a measure of entanglement. Operationally, EOF provides a computable bound on the entanglement cost~\cite{Bennett1996Mixed}, which quantifies the optimal state conversion rate of diluting maximally entangled states into the desired states under local operations and classical communication. Theoretically,
\begin{equation}
    E_{\rm F}(\rho_{AB})\geq\frac{S_\alpha-2\alpha}{2\sqrt{1+\alpha^2}-2\alpha},
    \label{eq:EOF_est_DI}
\end{equation}
where $\rho_{AB}$ is the underlying state shared by Alice and Bob. 

Another entanglement measure we aim to quantify is the negativity of entanglement. This measure is defined in terms of the violation of the positive partial transpose criteria~\cite{Vidal2002Computable}, and the corresponding bound is:
\begin{equation}
\label{eq:negativity_est_DI}
        \mathcal{N}(\rho_{AB})\geq \frac{S_\alpha-2\alpha}{4(\sqrt{1+\alpha^2}-\alpha)}.
\end{equation}

As for incompatibility, we use the effective overlap to measure it. This measure is state-dependent and can be tested experimentally~\cite{Tomamichel2013The}. In the standard CHSH case, the quantity is effectively upper-bounded by:
\begin{equation}
\label{eq:analytical_overlap}
    c^*(\rho_A,A_0,A_1)\leq \frac{1}{2}+\frac{S}{8}\sqrt{8-S^2},
\end{equation}
wherein the incompatibility is thus lower-bounded by $\min\{c^*,1-c^*\}$. One can find more explanation about these measures and multi-$\alpha$ incompatibility estimation results in Appendix Sec.~A.

\begin{figure*}[htbp]
\centering
\includegraphics[width =0.9\textwidth]{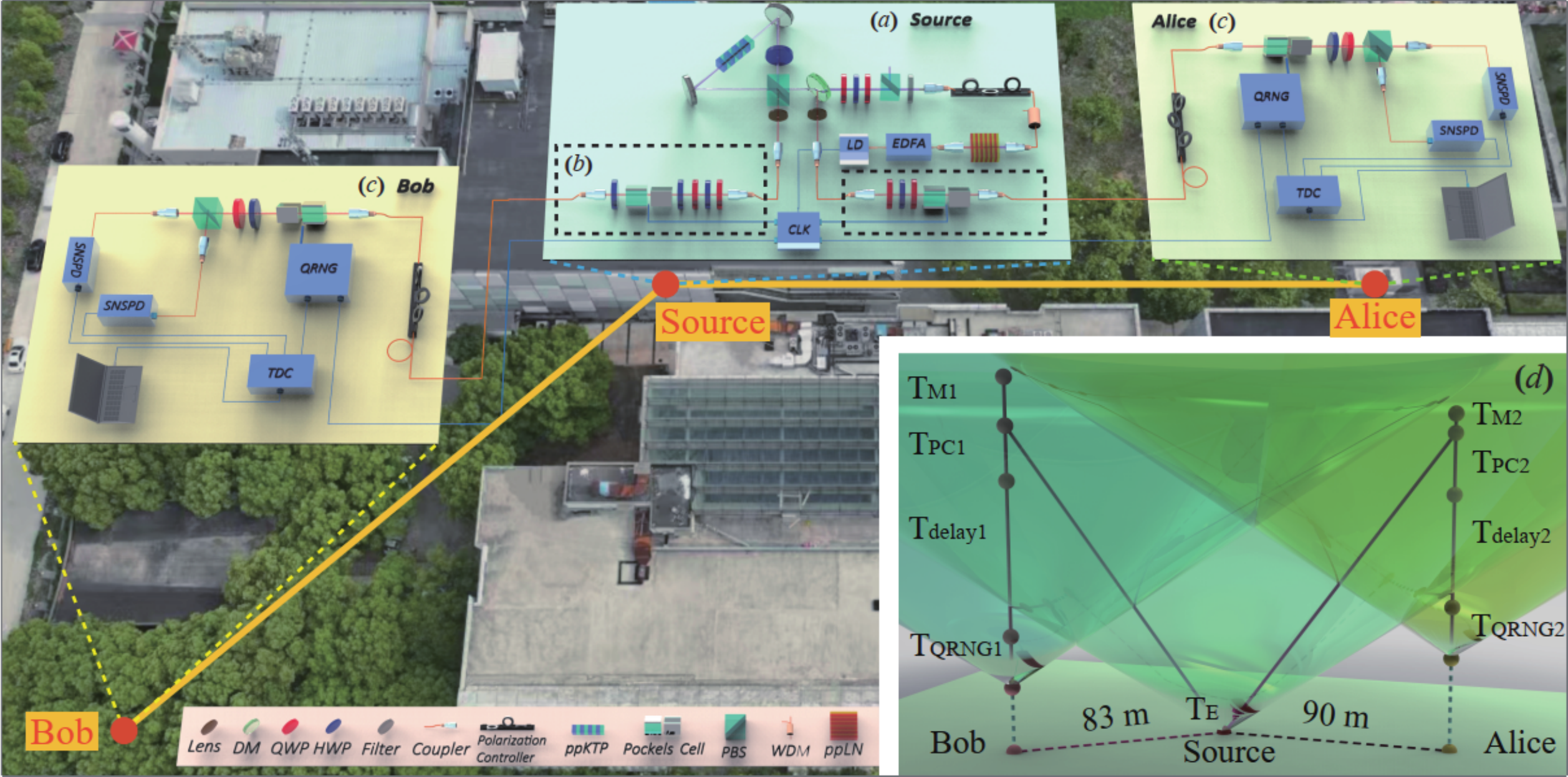}
\caption{Schematics of the experiment setup. (a) Preparation of entangled photon pairs. (b) Preparation of the Bell diagonal state in Eq.~\eqref{eq:Bell diagnal state}. (c) Single photon polarization state measurements for Alice and Bob, respectively, consist of a polarization controller (PC), a Pockels cell, a half-wave plate (HWP),  a polarizing beam splitter (PBS), and finally a superconducting nanowire single-photon detector (SNSPD). For the DI entanglement and measurement incompatibility evaluation experiment, there is only one detector on the transmission port of the PBS; for the interplay experiment, there are two detectors on both the transmission and reflection ports for post-selection. 
(d)The space-time correlation of the experiment. $T_\text{E} = 10$ ns is the time needed to create entangled photon pairs. $T_\text{QRNG1,2}$ represents the duration for generating random bits to switch Pockels cells. $T_\text{delay1,2}$ is the interval between random bit generation and delivery to the Pockels cells. $T_\text{PC1,2}$ indicates the delay for Pockels cells to prepare for state measurements after receiving the random bits. $T_\text{M1,2}$ denotes the duration for SNSPD to produce electronic signals. The details can be seen in Appendix Sec.~C.}
\label{fig:setup}
\end{figure*}
The experimental setup is shown in Fig.~\ref{fig:setup}. The polarization-entangled 1560 nm photon pairs are created by spontaneous parameter down conversion (SPDC) in the Sagnac loop from a 780 nm pump laser with a pulse width of 10 ns and a repetition rate of 200 kHz. Then, the photon pairs are transmitted to Alice's and Bob's laboratories separately through fiber channels for independent measurements. To close the \emph{locality loophole}, we separate Alice's and Bob's measurement stations far enough apart and apply fast measurements and precise synchronization, as shown in Fig.~\ref{fig:setup}(d). We designate the detection event on the transmission path of the PBS as $a(b)=1$; otherwise, $a(b)=-1$ for each trial. The overall detection efficiencies are $82.8\%\pm 0.2\%$ for Alice and $82.7\%\pm 0.2\%$ for Bob, respectively, which are high enough to surpass the threshold to close the \emph{detection loophole}.

\begin{figure}[h]
\centering
\includegraphics[scale=1.1]{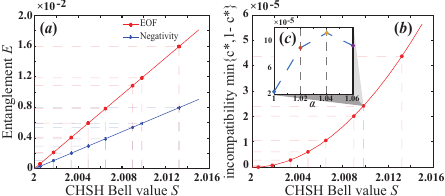}
\caption{(a) Diagram of EOF and negativity estimation results in the DI experiment. We plot the theoretical estimated results of EOF and negativity with red and blue solid lines, respectively. We mark the experimental estimated results of EOF and negativity with red and blue dots on each line, respectively. (b) Diagram of incompatibility estimation results in the DI experiment. We plot the theoretical estimated results of $\min\{c^*,1-c^*\}$ with a red solid line and mark the experimental estimated results with red dots. 
(c) Subdiagram of (b), illustrating the incompatibility estimation results with discrete multiple $\alpha$ using the DI experiment statistics of $S=2.0098$. Among the choices of $\alpha$, we can see the optimal estimation result of $S=2.0098$ statistic is obtained at the CHSH-type expression with $\alpha=1.04$, estimating $1.12\times 10^{-4}$ incompatibility of Alice's side.}
\label{fig:DI}
\end{figure}

After $5.76\times10^9$ trials of experiment, the final quantity $S_\alpha$ of interest, defined by Eq.~\eqref{eq:CHSH_type}, can be calculated from:
\begin{equation}
\label{eq:alpha_CHSH}
    \langle\hat{A}_{x}\otimes\hat{B}_{y}\rangle=\frac{N_{{}-1,{}-1|xy}-N_{{}-1,1|xy}-N_{1,{}-1|xy}+N_{1,1|xy}}{N_{xy}},
\end{equation}
where, $N_{xy}$ is the number of trials with inputs $x$ and $y$. And $N_{ab|xy}$ is the number of correlated events with outputs $a$ and $b$, given $x$ and $y$. We obtain eight distinct CHSH values $S$. In Fig.~\ref{fig:DI}(a), we quantify the entanglement of the underlying states of these eight results using EOF and negativity. For instance, upon reaching $S = 2.0132$, we certify the EOF in the underlying state to be 0.0159. More entanglement estimation results and $P$ values of CHSH violation $S$ are provided in Appendix Sec.~C.

Next, we obtain the quantification results for incompatibility using $c^*$. The estimation results are depicted in Fig.~\ref{fig:DI}(b). Additionally, by varying the parameter $\alpha$ when $S=2.0098$, we observe that for specific 
$\alpha>1$, a more precise estimation can be achieved via the CHSH-type expression given in Eq.~\eqref{eq:CHSH_type}. The estimation via multiple values of $\alpha$ is detailed in Appendix Sec.~B. In Fig.~\ref{fig:DI}(c), the optimal estimation of incompatibility, $1.12\times 10^{-4}$, is achieved at $\alpha=1.04$ among the provided options. 
\\
\\
\mysection{Interplay among three quantum resources}
In the previous part, by optimizing the entanglement, we obtain a monotonic relationship between nonlocality and incompatibility, as shown in Eq.~\eqref{eq:analytical_overlap}. Here, we extend our previous findings by investigating a more general interplay among these three pivotal quantities in quantum physics. Employing the control variable method, we held the level of entanglement constant to observe the relationship between nonlocality and incompatibility. To simplify the discussion, we assume that (1) the underlying system is a pair of qubits and (2) the measurement operators are qubit observables. 

In our experimental demonstration, we maintain a fixed amount of  entanglement in the Bell-diagonal state,
\begin{equation}
    \rho_\lambda=\lambda_1|\Psi^+\rangle\langle\Psi^+|+\lambda_2|\Psi^-\rangle\langle\Psi^-|+\lambda_3|\Phi^+\rangle\langle\Phi^+|+\lambda_4|\Phi^-\rangle\langle\Phi^-|,
\label{eq:Bell diagnal state}
\end{equation}
where $\ket{\Phi^\pm}=(\ket{00}\pm\ket{11})/\sqrt{2},\ket{\Psi^\pm}=(\ket{01}\pm\ket{10})/\sqrt{2}$ are four Bell bases. The angle difference $\theta$ between two measurements at Bob are varied to change the incompatibility. In this case, to illustrate the interplay, we take concurrence and one-way distillable entanglement (ODE) as the target entanglement measures, since the former captures EOF (negativity) and the latter captures entanglement entropy for the Bell-diagonal state, respectively. In our experimental setup, we select states with the concurrence $C=0.4$ and ODE $E_D^\rightarrow=0.1$, $E_D^\rightarrow=0.2$, and we analyze the $\alpha$-CHSH value for $\alpha=1$ and $\alpha=1.5$ separately. The specific states chosen for this analysis are detailed in the TABLE~\ref{tab:state}.

\begin{table}[]
    \centering
    \caption{The states for analysis of the interplay among entanglement, nonlocality, and measurement incompatibility.} 
    \label{tab:state}
    \begin{tabular}{c|c |c c c|c c c}
    \hline
    \hline
      ~&$C=0.4$&\multicolumn{3}{c}{$E_D^\rightarrow=0.2$}&\multicolumn{3}{c}{$E_D^\rightarrow=0.1$}  \\
      \hline
      ~&state 0&state 1&state 2&state 3&state 4&state 5&state 6\\
      $\lambda_1$&0.7&0.788&0.832&0.847&0.797&0.765&0.712 \\
      $\lambda_2$&0.3&0.203&0.132&0.079&0.163&0.215&0.284\\
      $\lambda_3$&0&0.006&0.031&0.068&0.033&0.016&0.003\\
      $\lambda_4$&0&0.002&0.005&0.006&0.007&0.004&0.001\\
      \hline
    \hline
    \end{tabular}

\end{table}

To prepare the four Bell bases, we utilize a combination of waveplates and a Pockels cell to conduct local operations along the paths to Alice and Bob. 
On the measurement side, we combine a PC, a HWP, and a Pockels cell to control the measurement settings (see Methods). 
Finally, we employ post-selection to process the obtained data. Given the utilization of SPDC sources, the Poisson distribution~\cite{Caprara2015Challenging} governing the photon count contributes a large number of zero-photon states. To filter out the data corresponding to zero-photon events and guarantee that the experimental data originates from the entangled states we have generated, we designate the detection event on the transmission path of the PBS as $a(b)=1$, and the event on the reflection path as $a(b)=-1$ for each trial.

\begin{figure}[htbp]
\centering
\includegraphics[width=8.5cm]{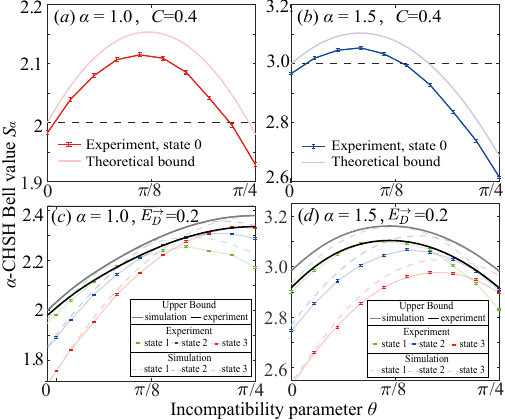}
\caption{The interplay between nonlocality and incompatibility under a fixed level of entanglement. Each subfigure presents the primary outcome through a dark-colored line: the experimental trajectory of the maximum $\alpha$-CHSH value achievable at a specific entanglement level while varying incompatibility. In subfigures (a) and (b), the maximum $\alpha$-CHSH value is derived using a constant state, state 0, with $C=0.4$, such that the red and blue line segments (with error bars indicating a standard deviation) in (a) and (b) depict the interplay. The faint solid line represents the theoretical upper bound of the $\alpha$-CHSH value. Conversely, in subfigures (c) and (d), the maximum $\alpha$-CHSH value is obtained from distinct states with a fixed ODE of $E_D^{\rightarrow}=0.2$, resulting in the interplay trajectory depicted by the convex hull of states 1, 2, and 3, shown by the faint line segments (with error bars). The upper bounds simulated for these states are shown as dashed lines, while the theoretical interplay trajectory is illustrated by a faint gray line.}
\label{fig:Interplay_con}
\end{figure}

The trajectory is illustrated in Fig.~\ref{fig:Interplay_con}, showcasing the experimental results of the observed mixed entangled states, state 0 with $C=0.4$ and states 1,2,3 with $E_D^{\rightarrow}=0.2$. Our experimental findings reveal a counterintuitive correlation: while entanglement remains a constant, as incompatibility increases, the $\alpha$-CHSH value initially ascends and subsequently descends. This behavior is evident in both entanglement measures for $\alpha=1.5$, also in concurrence for $\alpha=1$. It witnesses the generic interplay among nonlocality, entanglement, and incompatibility. Simultaneously, our experimental trajectory fits in perfectly with our theoretical simulation trajectory, which verifies the solidness of theoretical results.
\\
\\
\mysection{Conclusion and discussion}
In conclusion, we have demonstrated a device-independent method to quantify the entanglement and measurement incompatibility of a quantum system. By successfully closing various loopholes in the optical system, we show that a loophole-free Bell test setup can quantify a significant amount of entanglement based on EOF and negativity. Also, we optimize the estimation of measurement incompatibility by changing the tilted parameter $\alpha$ of the generalized CHSH inequality.

With the quantification of entanglement through nonlocality, we delve into the interplay among nonlocality, entanglement, and measurement incompatibility. Our experimental findings reveal a counterintuitive phenomenon within quantum mechanics: given a certain level of entanglement, an increase in measurement incompatibility unexpectedly leads to a decrease in Bell nonlocality. This runs counter to the conventional expectation that greater quantumness should enhance violations of locality. This surprising observation challenges established notions and underscores the intricate subtleties of quantum theory.

Building upon this discovery, we present that, given a certain amount of entanglement, two distinct levels of incompatibility can yield equivalent nonlocality. Consequently, we propose that the interplay among three quantities offers promising guidance when optimizing resource allocation strategies based on Bell nonlocality. Our experiment paves the way for practical applications of quantum resource theory.

To reveal the interplay among the three quantum resources experimentally, we fix the entanglement of Bell-diagonal states. In our method for preparing Bell-diagonal states, we set the number of occurrences of each Bell basis per cycle as $\lambda_i n$ (with $n=1000$) to generate a statistical target Bell-diagonal state. In the context of a physical mixture, each cycle is characterized by a single pulse ($n=1$), resulting in the probabilistic occurrence of the four Bell bases. This stochastic nature can be effectively implemented using a random number generator.

There exists an interesting finding in Fig.~\ref{fig:Interplay_con}. We found that there are slight horizontal and vertical shifts between the experimental and theoretical trajectories. The measurement deviations cause the horizontal shift, and independently, state fidelity deviations presumed to be solely due to white noise cause the vertical shift. This observation enables us to precisely calibrate the measurement devices and the state fidelity separately. The calibration procedure and enhancements to our system are detailed in Appendix  Sec.~C.
\\
\\
\mysection{Method}

\begin{figure}[htbp]
\centering
\includegraphics[width=0.5\linewidth]{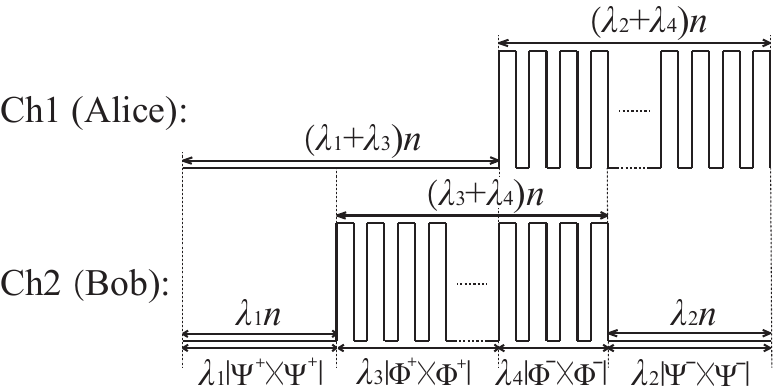}
\caption{Time sequence for producing target Bell-diagonal states. To prepare $\lambda_4|\Phi^-\rangle\langle\Phi^-|$ component, operations are required on both Alice and Bob's ends, necessitating $\lambda_4n$ ON pulses for both Channels 1 and 2. Conversely, for preparing the $\lambda_2|\Psi^-\rangle\langle\Psi^-|$ state, solely Alice's side needs
the $\sigma_z$ operation, resulting in $\lambda_2$ ON pulses for Channel 1 and OFF pulses for Channel 2.}
\label{fig:mixed state}
\end{figure}

\noindent\textbf{Preparation of mixed state.} To prepare four Bell states, we utilize a combination of waveplates and a Pockels cell to conduct local operations along the paths to Alice and Bob. The transformation between these four bases can be achieved through the following operations:
\begin{equation}
    \begin{split}
     &|\Psi^-\rangle=\sigma_z\otimes~I\cdot|\Psi^+\rangle, \\&|\Phi^+\rangle=I~\otimes\sigma_x\cdot|\Psi^+\rangle, 
     \\&|\Phi^-\rangle=\sigma_z\otimes\sigma_x\cdot|\Psi^+\rangle
\end{split}
\end{equation}
where $\sigma_x,\sigma_y$, and $\sigma_z$ are the Pauli operators. In our experimental setup depicted in Fig.~\ref{fig:setup}(b), the entangled states directly generated from the Sagnac loop are $\ket{\Psi^+}$. When the high voltage is applied to the Pockels cell, the $\sigma_z$ ($\sigma_x$) operation is conducted along the path to Alice's (Bob's) side; otherwise, the operation is the identity operator $I$. These enable us to fast switch among the four Bell bases. We use a multichannel pulse generator, which is synchronized to the system clock, to produce triggers for high-voltage power. We regard $n$ pulses as a cycle. At the beginning of each cycle, Channel 1 (Ch1) for Alice sequentially generates $(\lambda_2+\lambda_4)n$ ON pulses to drive the high-voltage power, followed by OFF pulses till this cycle ends. Concurrently in the same cycle, Channel 2 (Ch2) for Bob waits for $\lambda_2n$ OFF pulses, then produces $(\lambda_3+\lambda_4)n$ ON pulses before waiting for the cycle ends (as shown in Fig.~\ref{fig:mixed state}). We take $n=1000$ in the experiment. After repeating $1.2\times10^5$ cycles, the target state as defined in Eq.~\eqref{eq:Bell diagnal state} can be statistically obtained, with a fidelity exceeding $99.10\%$.

\noindent\textbf{Rotation of measurement settings.} On the measurement side, PC rotates the polarization of incident light by an angle $\gamma$, and HWP is set at angle $\omega$. Measurement $A_1$ is performed when high voltage is applied to the Pockels cell, otherwise, measurement $A_0$ is performed. Thus the relation between the angle of measurement settings $\theta_{A_0}, \theta_{A_1}$ ($\theta_{B_0}, \theta_{B_1}$) and PC and HWP is:
\begin{equation}
\begin{split}
      &\gamma = 45^\circ - \frac{\theta_{A_0}(\theta_{B_0})+\theta_{A_1}(\theta_{B_1})}{2}\\
      &\omega = 22.5^\circ + \frac{\theta_{A_0}(\theta_{B_0})-\theta_{A_1}(\theta_{B_1})}{4}
      \end{split}
    \label{eq:measurement}
\end{equation}
where $A_0=\cos(2\theta_{A_0})\sigma_z+\sin(2\theta_{A_0})\sigma_x$, The PC will only affect the sum of two measurement settings, and HWP will affect the difference between them. In the experiment, we hold the measurement of Alice to be $A_0=X$ and $A_1=Z$ while changing the measurements of Bob every ten minutes. We leave the PC of Bob unchanged and rotate the HWP from $22.5^\circ$ to $11.25^\circ$ in 100 minutes with a step of $-1.25^\circ$.
\\
\\
\mysection{Data availability}
All data that support the plots within this
paper and other findings of this study are available from the corresponding authors upon reasonable request.
\\
\mysection{Code availability}
All relevant codes or algorithms are available from the corresponding authors upon
reasonable request.
\\
\mysection{Acknowledge}
Hai-Hao Dong, Yuwei Zhu, and Su-Yi Cheng contributed equally to this work. This work has been supported by the National Natural Science Foundation of China (Grants No. T2125010, No. 12174216), the Shanghai Municipal Science and Technology Major Project (Grants No. 2019SHZDZX01), the Anhui Initiative in Quantum Information Technologies (Grant No. AHY010300), the Innovation Program for Quantum Science and Technology (Grants No. 2021ZD0300800, No. 2023ZD0300100, No. 2021ZD0300804 and No. 2021ZD0300702.)

\begin{appendix}
\section{Details behind the theoretical results}
\subsection{Entanglement measures}
Consider a bipartite state $\rho_{AB}\in\mathcal{D}(\mathcal{H}_A\otimes\mathcal{H}_B)$ acting on the associated Hilbert space. The first entanglement measure we examine is the entanglement of formation (EOF), denoted as $E_{\mathrm{F}}(\rho_{AB})$~\cite{Bennett1996Mixed}. For pure states, the EOF is equivalent to the entanglement entropy, given by $E_{\mathrm{F}}(\ket{\phi}_{AB})=H(\rho_A)=H(\rho_B)$, where $\rho_A$ and $\rho_B$ are the reduced states of systems $A$ and $B$ obtained by tracing out the other subsystem, and $H(\cdot)$ denotes the von Neumann entropy. Generalizing to mixed states, the EOF is defined using a convex-roof construction 
\begin{equation}
    E_{\mathrm{F}}(\rho_{AB})=\min_{\{p_i,\ket{\phi}_i\}_i}\sum_{i}p_i E_{\mathrm{F}}(\ket{\phi_i}_{AB}),
    \label{EOF}
\end{equation}
where the optimization is carried out over all possible pure-state decompositions. In the realm of two-qubit states, the EOF simplifies to the expression in~\cite{hill1997entanglement}
\begin{equation}
    E_{\mathrm{F}}(\rho_{AB})=h\left(\frac{1+\sqrt{1-C^2(\rho_{AB})}}{2}\right),
\label{eof}
\end{equation}
where $h(p)=-p\log p-(1-p)\log (1-p)$ is the binary entropy function for $p\in[0,1]$, and $C(\rho_{AB})$ is the concurrence of $\rho_{AB}$, a useful entanglement monotone~\cite{hill1997entanglement,rungta2001universal}.

Another crucial entanglement measure we consider is the negativity of entanglement, defined in terms of the violation of the positive partial transpose (PPT) criteria, as given in~\cite{Vidal2002Computable},
\begin{equation}
\mathcal{N}(\rho_{AB})=\frac{\|\rho_{AB}^{\mathrm{T}_{A}}\|_1-1}{2}=\sum_{\lambda_i(\rho_{AB}^{\mathrm{T}_{A}})<0}|\lambda_i(\rho_{AB}^{\mathrm{T}_{A}})|,
\label{negativity}
\end{equation}
where $(\cdot)^{\mathrm{T}_{A}}$ is the partial trace operation on subsystem $A$ on the computational basis and $\|\cdot\|_1$ is the trace norm of a matrix. In the second equality of Eq.~\eqref{negativity}, $\lambda_i(\cdot)$ represents the eigenvalues of a matrix. Additionally, the logarithm of negativity, $E_{\mathcal{N}}(\rho_{AB})=\log\|\rho_{AB}^{\mathrm{T}_{A}}\|_1$
, provides an upper bound on distillable entanglement and is at least as large as the negative conditional entropy of entanglement~\cite{guhne2009entanglement}. Another notable issue is that the negativity of entanglement is closely related to the concurrence for a pair of qubits considering the underlying state to be a mixture of two-qubit Bell states~\cite{zhu2023entanglement}.

Another entanglement measure we consider in the interplay section is the one-way distillable entanglement, which quantifies the maximum rate of distilling maximally entangled states through one-way LOCC operations between two users. In the Shannon limit, obtained by considering infinitely many independent and identical (i.i.d.) copies of the quantum state, the average one-way distillable entanglement can be determined using the negative conditional entropy~\cite{wilde2017converse} (refer to Sec.~VIB in that work),
\begin{equation}
    E_D^\rightarrow(\rho_{AB}) = -H(A|B)_{\rho},
\end{equation}
where $H(A|B)_{\rho} = H(\rho_{AB}) - H(\rho_B)$. This result extends the earlier discovery in Ref.~\cite{Bennett1996Mixed}, which established that the one-way distillable entanglement in the Shannon limit is $1 - H(\rho_{AB})$ when $\rho_{AB}$ is a mixture of Bell states (see Sec.~IIIB3 in that reference).

The theoretical estimation results of the EOF and negativity based on the $\alpha$-CHSH Bell value can be found in Theorem 2 and Theorem 3 of Ref.~\cite{zhu2023entanglement}, respectively.

\subsection{Incompatibility measure}
\label{appsc:incompatibility}

The measure we use to quantify the incompatibility between two local measurements in this context is the effective overlap~\cite{Tomamichel2013The}. Consider Alice and Bob share the quantum state $\rho_{AB}$ acting on the Hilbert space $\mathcal{H}_A\otimes\mathcal{H}_B$. Let $A_i$ and $B_j$ represent binary outcome measurements for Alice and Bob, respectively, given inputs $i$ and $j$. The incompatibility between observables $A_0$ and $A_1$ on state $\rho_A$ can be assessed using the effective overlap, denoted as $c^*(\rho_A, A_0, A_1)$, as defined in Ref.~\cite{Tomamichel2013The}. In a Bell test, the violation of Bell inequality necessarily indicates entanglement and local measurement incompatibility. In the simplest Bell test, the CHSH Bell test, when the Bell value is $S$, the effective overlap between local measurements on either side in the underlying system is theoretically upper-bounded by

\begin{equation}
\label{eq:analytical_overlap}
    c^*(\rho_A,A_0,A_1) \leq \frac{1}{2} + \frac{S}{8}\sqrt{8-S^2}.
\end{equation}

In essence, the incompatibility between $A_0$ and $A_1$ is lower-bounded by a function of $S$. By extending the consideration to a wider range of CHSH Bell tests with an additional parameter $\alpha$, the incompatibility estimation problem can be solved by the optimization problem,

\begin{equation}
    \begin{split}
     I_{\min}& =\underset{\rho_{AB},A_0,A_1,B_0,B_1}{\text{min}} \min\{c^*,1-c^*\}, \\
    \text{s.t.}\quad
    S &=\Tr[\rho_{AB}\left(\alpha\hat{A_0}\otimes \hat{B_0}+\alpha \hat{A}_0\otimes \hat{B}_1+\hat{A}_1\otimes \hat{B}_0-\hat{A}_1\otimes \hat{B}_1\right)],\\
    \rho_{AB} & \geq 0;\\
    \Tr(\rho_{AB}) & =1.
    \end{split}
    \label{eq:optm_overlap}
\end{equation}

\begin{figure}[htbp]
\centering
\includegraphics[width=12cm]{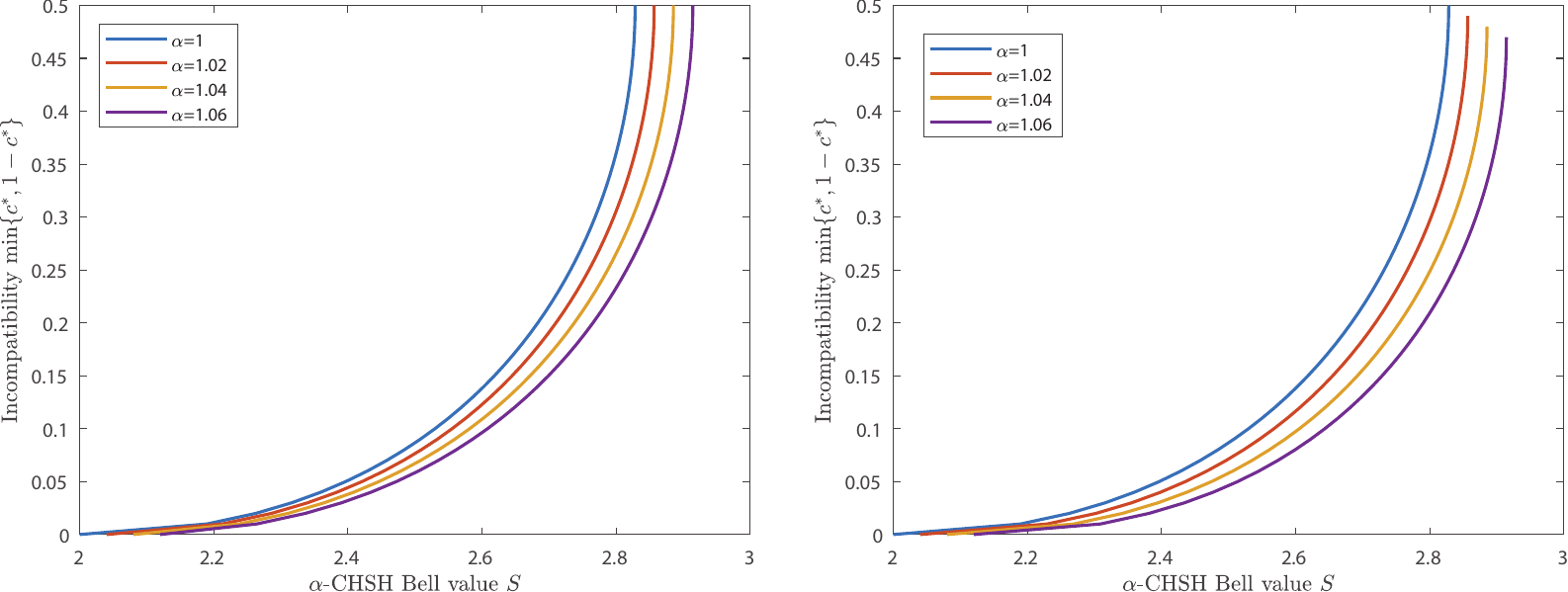}
\caption{Illustration of CHSH-type Bell value as a function of effective overlap. In two figures, we present the estimation results when $\alpha=1,1.02,1.04$ and $1.06$. On the left is the local incompatibility estimation result of Alice's side, while on the right is Bob's side.}
\label{fig:overlap_A_B_multi_alpha}
\end{figure}

Here, $S$ represents the corresponding $\alpha$-CHSH Bell value. When $\alpha=1$, the solution to Eq.~\eqref{eq:optm_overlap} degenerate to the analytical result in Eq.~\eqref{eq:analytical_overlap}. Numerical results for different $\alpha$ values can be found in Fig.~\ref{fig:overlap_A_B_multi_alpha}.

\section{Theoretical part of interplay}

In this section of the Appendix, we delve into the interplay relationship between nonlocality, entanglement, and measurement incompatibility. To illustrate it, we focus on an optimization problem defined as follows:
\begin{equation}
\label{eq:interplay_optm_fixed_E}
    \begin{split}
     &\max_{\rho_{AB}} \Tr\left(\rho_{AB}\hat{S}_{\alpha}\right) , \\
    \text{s.t.}\quad
    E(\rho_{AB}) &= E,\\
   \hat{A}_0 &= \sigma_z,\\
   \hat{A}_1 &= \sigma_x,\\
    \hat{B}_0 &=\cos\theta\sigma_z+\sin\theta\sigma_x,\\
    \hat{B}_1 &=\cos\theta\sigma_z-\sin\theta\sigma_x,\\
    \rho_{AB}&\geq 0, \\
    \rho_{AB}&\in\mathcal{D}(\mathcal{H}_2\otimes\mathcal{H}_2), \\
    \Tr(\rho_{AB})&=1.
    \end{split}
\end{equation}
The input state $\rho_{AB}$ for qubit pairs exhibits entanglement with a specified value $E$, while Alice's measurements are fixed as projective-valued measurements of $\sigma_z$ and $\sigma_x$ in Pauli operators, designed to be maximally incompatible. Bob's measurements are parameterized by $\theta\in[0,\pi/4]$. When $\theta=0$ and $\pi/2$, Bob's local observables commute, but at $\theta=\pi/4$, they reach maximal incompatibility. The setup indicates an $\alpha$-CHSH Bell value $S$. By solving Eq.~\eqref{eq:interplay_optm_fixed_E}, it was revealed that, for $\alpha > 1$, the relationship between nonlocality and measurement incompatibility is not a straightforward monotone increase when a fixed level of entanglement is maintained. Specifically, the investigation showed that decreasing measurement incompatibility can paradoxically lead to an increasing nonlocality. In Ref.~\cite{zhu2023entanglement}, this intriguing interplay is originally explored from an alternative perspective, where the Bell nonlocality is settled, and the optimization goal shifts to minimizing the required entanglement. This alternative viewpoint plays the role as a duality to the optimization problem outlined in Eq.~\eqref{eq:interplay_optm_fixed_E}.

\section{experimental details}
\subsection{Quantum state characterization}
We perform the state tomography measurement on the non-maximally entangled state and mixed state. We use the maximum likelihood estimation of density matrices to calculated the fidelity to avoid problem of experimental inaccuracies and statistical fluctuation of coincidence counts.

\begin{figure}[htbp]
\centering
\includegraphics[width =1\textwidth]{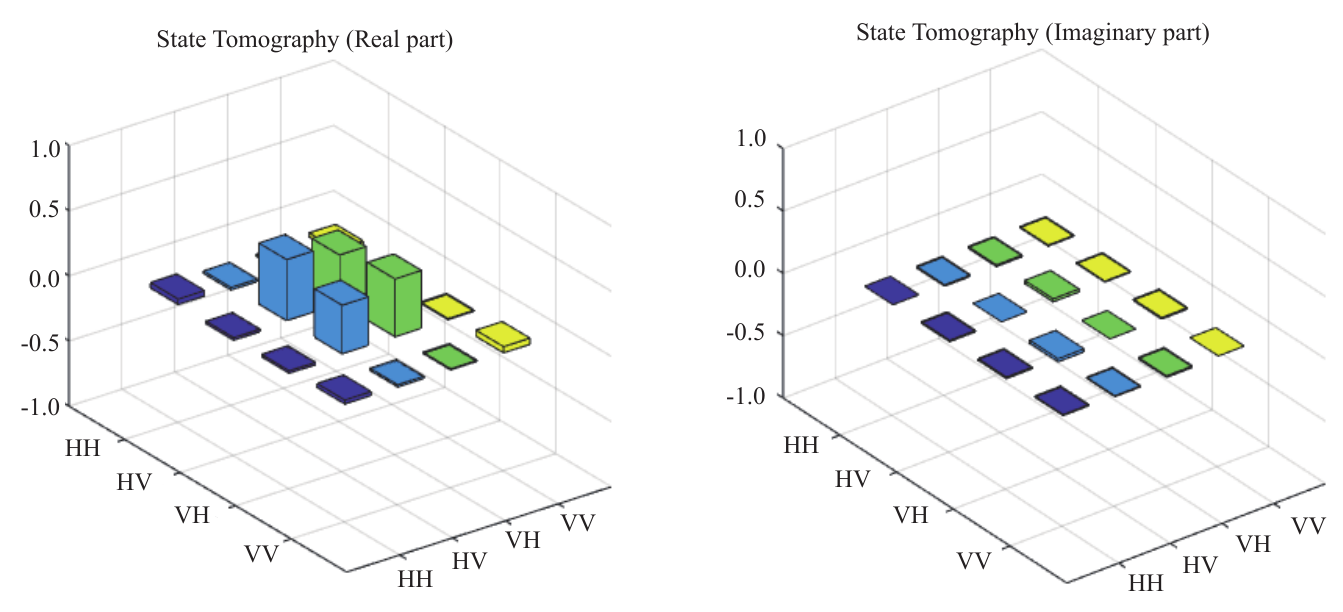}
\caption{(color online) Tomography of the prepared quantum state. The real and imaginary parts are shown in (a) and (b), respectively.}\label{tomo}
\end{figure}

We generate a formula for an explicitly ‘‘physical’’ density
matrix, i.e., a matrix that has the three important properties
of normalization, Hermiticity, and positivity. This matrix will
be a function of 16 real variables with $t=\{t_1 ,t_2 , . . . ,t_{16}\}$ and is denoted as $\rho_p(t)$. For any matrix that can be written in the form $G = T^\dag T$ must be non-negative definite. The explicitly ‘‘physical’’ density matrix $\rho_p$ is given by the formula
\begin{equation}
    \rho_p = T^\dag(t)T(t)/\text{Tr}\{T^\dag(t)T(t)\}
\end{equation}
and it is convenient to choose a tridiagonal form for $T$ :
\begin{equation}
    T(t) = \left[ {\begin{array}{cccc}t_{1} & 0 & 0 & 0 
    \\t_{5}+it_6 & t_{2} & 0 & 0 
    \\t_{11}+it_{12} & t_{7}+it_8 & t_{3} & 0 
    \\t_{15}+it_{16} & t_{13}+it_{14} & t_{9}+it_{10} & t_{4}
    \\\end{array} } \right]
\end{equation}
The measurement data consists of a set of coincidence
counts $n_\mu$ whose expected value is $\overline{n_\mu}=N\langle\phi_\mu|\rho|\phi_\mu\rangle$. Here $\rho$ is the prepared quantum state. In our experiment $\mu = 1,2,...,36$, $|\phi_\mu\rangle\langle\phi_\mu|$ is the operator of the projection measurement of the two-photon state, and for each photon we measured in the bases $|H\rangle,|V\rangle,|+\rangle,|-\rangle,|R\rangle,|L\rangle$, there are 36 projection measurements for $\rho$, $\phi_\mu\in{|HH\rangle,|HV\rangle,......|RL\rangle,|RR\rangle}$. Assuming that the noise on these coincidence measurements has a Gaussian probability distribution. Thus the probability of obtaining a set of 36 counts ${n_1 ,n_2 , . . . n_{36}}$ is 
\begin{equation}
    P(n_1 ,n_2 , . . . n_{36}) = \frac{1}{N_{norm}}\prod\limits_{\mu=1}^{36}\exp[-\frac{(n_{\mu}-\overline{n_\mu})^2}{2\sigma^2_\mu}]
\end{equation}
where $\sigma_\mu$ is the standard deviation for the $n$-th coincidence measurement (given approximately by $\sqrt{\overline{n_\mu}}$) and $N_{norm}$ is the normalization constant. For our physical density matrix $\rho_p$, the number of counts expected for the $n$-th measurement is
\begin{equation}
    \overline{n_\mu}(t_1,t_2,...,t_{16}) = N\langle\phi_\mu|\rho_p(t_1,t_2,...,t_{16})|\phi_\mu\rangle
\end{equation}
Thus the likelihood that the matrix $\rho_p(t_1,t_2,...,t_{16})$ could
produce the measured data $n_1 ,n_2 , . . . ,n_{36}$ is
\begin{equation}
    P(n_1 ,n_2 , . . . n_{36}) = \frac{1}{N_{norm}}\prod\limits_{\mu=1}^{36}\exp[-\frac{(N\langle\phi_\mu|\rho_p(t_1,t_2,...,t_{16})|\phi_\mu\rangle-n_\mu)^2}{2N\langle\phi_\mu|\rho_p(t_1,t_2,...,t_{16})|\phi_\mu\rangle}]
\end{equation}
Rather than to find the maximum value of $P(t_1 ,t_2 , . . . ,t_{16})$, it simplifies things somewhat to find the maximum of its logarithm (which is mathematically equivalent). Thus the optimization problem reduces to finding the minimum of the following function:
\begin{equation}
    L(t_1 ,t_2 , . . . t_{16}) = \sum_{\mu=1}^{36}\frac{(N\langle\phi_\mu|\rho_p(t_1,t_2,...,t_{16})|\phi_\mu\rangle-n_\mu)^2}{2N\langle\phi_\mu|\rho_p(t_1,t_2,...,t_{16})|\phi_\mu\rangle}
\end{equation}
This is the ‘‘likelihood’’ function that we employed in our numerical optimization routine.
The result of state 3 is shown in
Fig.~\ref{tomo}, the state fidelity is $99.49\%$. The coordinate axis labeled by HH, HV, VH and VV for the density operator$\rho$ can be written as:
\begin{equation}
\rho = \rho_{11}|HH\rangle\langle HH|+\rho_{12}|HH\rangle\langle HV|+...+\rho_{44}|VV\rangle\langle VV|
\end{equation}
where $\rho_{ij}$ if the elements of the density operator matrix. The main imperfections are attributed to the multi-photon components, imperfect optical elements, and imperfect spatial/spectral mode matching.

\subsection{Spacetime configuration of the experiment}\label{App:spacetime_details}

\begin{figure}[htbp]
\centering
\includegraphics[width =0.7\textwidth]{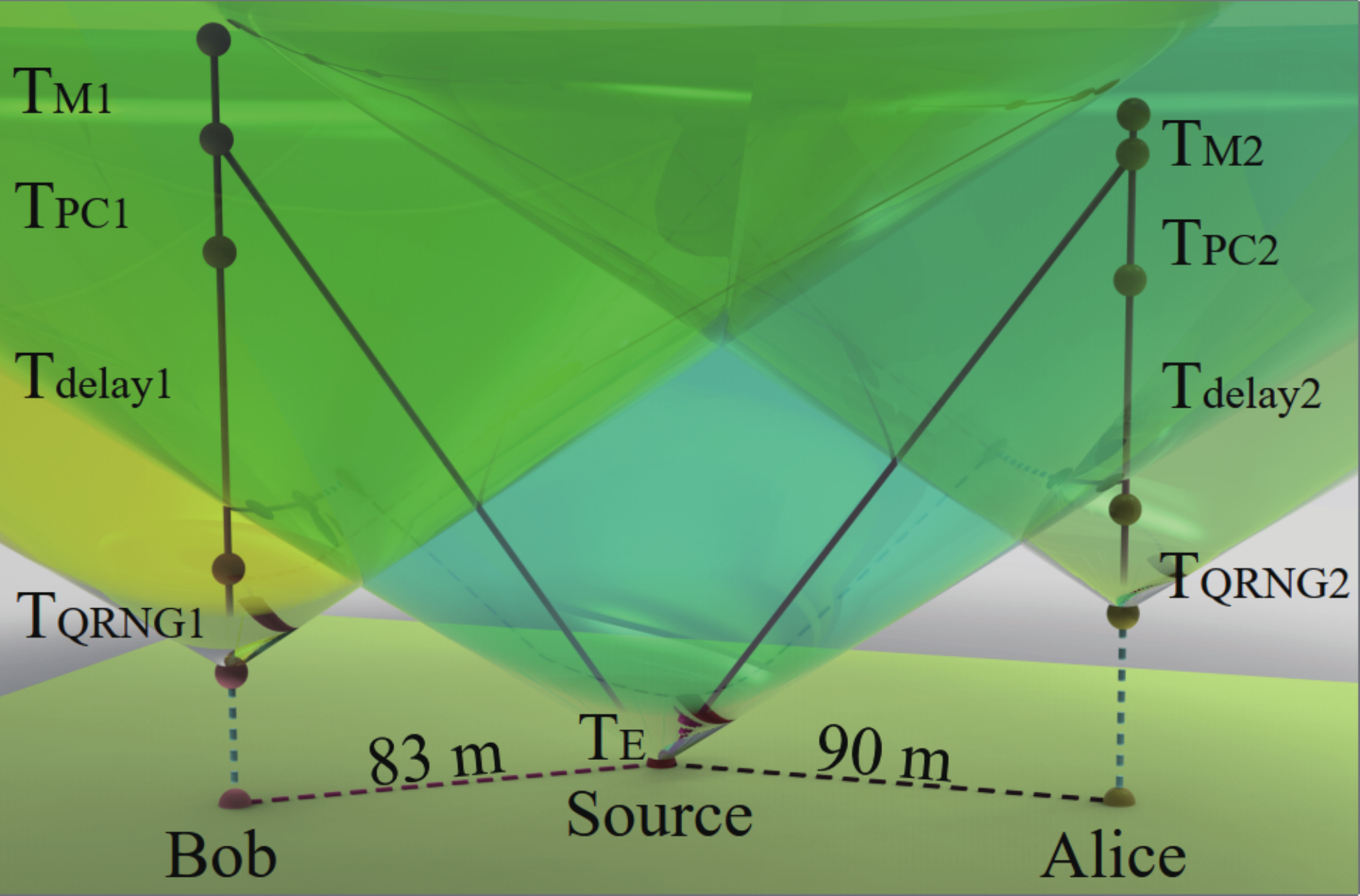}
\caption{Space-time diagram for the experimental events. $T_{E}=10 ~\text{ns}$ is the generation time of entangled photon pairs. $T_{QRNG1,2}$ are duration of times required to generate random bits to switch the Pockels cells. $T_{delay1,2}$ are the delay time between the random bits being generated and received by the Pockels cells. $T_{PC1,2}$ are the waiting times for the Pockels cells to perform state measurements after receiving the random bits. $T_{M1,2}$ are the times taken by the single photon detectors to output electronic signals. $T_{QRNG1} = T_{QRNG2} = 96 ~\text{ns}$, $T_{delay1} = 208 ~\text{ns}$, $T_{delay2} = 287 ~\text{ns}$, $T_{PC1} = 112 ~\text{ns}$,
$T_{PC2} = 100 ~\text{ns}$, $T_{M1} = 25~\text{ns}$ and $T_{M2} = 77~ \text{ns}$. Alice’s and Bob’s measurement stations are placed on opposite sides of the source at distances of 90 m and 83 m, respectively. The effective optical length between Alice’s (Bob’s) station and the source is 178 m (182 m). This arrangement ensures no signaling between relevant events in the experiment. The shaded areas are the future light cones for the source, Alice's and Bob's laboratories.}\label{spacelike}
\end{figure}

To rule out the locality loophole, space-like separation must be satisfied between measurement events at Alice and Bob’s measurement stations, i.e., the setting choice and measurement outcome at one station must be space-like separated from the events at the other station (see Fig.~\ref{spacelike}). Therefore, we then obtain

\begin{equation}
\left\{\begin{array}{l}
(|AB|) / c>T_{E}-\left(L_{SA}-L_{SB}\right) / c+T_{QRNG1}+T_{\text {Delay}1}+T_{PC1}+T_{M2} \\
(|AB|) / c>T_{E}+\left(L_{SA}-L_{SB}\right) / c+T_{QRNG2}+T_{\text {Delay}2}+T_{PC2}+T_{M1}
\end{array}\right.
\end{equation}
where $(|AB| = 163~\text{m})$ is the free space distance between Alice and Bob’s measurement station. $T_E = 10~\text{ns}$ is the generation time for photon pairs, which is mainly contributed by the $10~\text{ns}$ pump pulse duration. $L_{SA} = 178~\text{m}$ $(L_{SB} = 182~\text{m})$ is the effective optical path that is mainly contributed by the $122~\text{m} (124~\text{m})$ long fiber between the source and Alice's (Bob’s) measurement station. $T_{QRNG1} = T_{QRNG2} = 96~\text{ns}$ is the time elapsed for QRNG to generate a random bit. $ T_{Delay1} = 208~\text{ns}$ $(T_{Delay2} = 287~\text{ns})$ is the delay between the random numbers being generated and delivered to the Pockels cell. $ T_{PC1} = 112~\text{ns}$ $(T_{PC2} = 100~\text{ns})$ is the waiting time for the Pockels cell to be ready to perform state measurements after receiving the random numbers, including the internal delay of the Pockcels Cells $(62~\text{ns}, 50~\text{ns})$ and the time for Pockcels cell to stabilize before performing single photon polarization state projection after switching of $50~\text{ns}$. $T_{M1} = 25~\text{ns}$ $(T_{M2} = 77~\text{ns})$ is the time elapse for SNSPD to output an electronic signal, including the delay due to fiber and cable length. The electronic signals generated by SNSPD are considered clonable and thus represent a definite classical entity. Meanwhile, the photon involved in the entangled state is considered unclonable according to quantum physics. Therefore, we can reasonably assume that the photons collapse upon reaching the SNSPD, thereby concluding the measurement of the photon at the SNSPD. In that sense, the collapse-locality loophole, which concerns where the measurement outcomes eventually arise in space-time, was strongly tightened.

Measurement independence between entangled-pair creation events and setting choice events is satisfied by the following spacelike separation configuration:
\begin{equation}
\left\{\begin{array}{l}
|S A| / c>L_{S A} / c-T_{\text {Delay } 1}-T_{P C 1}, \\
|S B| / c>L_{S B} / c-T_{\text {Delay } 2}-T_{P C 2}.
\end{array}\right.
\end{equation}
where $|SA| = 90~\text{m}$ $(|SB| = 83~\text{m})$ is the free space distance between the entanglement source and Alice’s (Bob’s) measurement station. We estimate the fiber length by measuring the reflection, the single photon arrives at the SNSPD and generates an electronic response with high efficiency. However, with a small possibility, the photon is reflected by the SNSPD chip, travels to the source, gets polarization rotated in the Sagnac loop, and then travels back to SNSPD, making the second click. We measure the electronic cable length using a ruler. By subtracting the time for the signal passing through the fiber and cable, as well as the delay caused by the discriminator, we estimate the effective fiber length between the Pockels cell and the SNSPD chip. The measured fiber length, cable length, and discriminating time are summarized in Tab.~\ref{tab:length}.

\setlength{\tabcolsep}{8mm}
\begin{table}[ht]
    \centering
    \caption{The fiber distances between Source and Measurement.}
\begin{tabular}{c|ccc}
\hline
      & \textbf{Source-PC} & \textbf{PC-SNSPD} & \textbf{SNSPD-TDC} \\ \hline
\textbf{Alice} & 122 m     & 5 m     & 4 m       \\
\textbf{Bob}   & 124 m     & 15 m     & 12 m      \\ \hline
\end{tabular}
\label{tab:length}
\end{table}

\subsection{Quantification of entanglement and measurement incompatibility}
We obtain eight distinct CHSH values $S$. In Tab.~\ref{tab:result}, we quantify the entanglement of the underlying states of these eight results using EOF and negativity. We obtain the quantification result for incompatibility using $c^*$.
\begin{table}[h]
\centering
\caption{Device-independent quantification of entanglement and measurement incompatibility}
\setlength{\tabcolsep}{2mm}
\begin{tabular}{c|c c c c c c c c}
     \hline
     Bell violation $S$&2.0005&2.0017&2.0033&2.0049&2.0065&2.0089&2.0098 &2.0132\\
     \hline
     EOF&0.0006&0.0021&0.0041&0.0059&0.0078&0.0108&0.0118&0.0159\\
     Negativity&0.0003&0.0011&0.0020&0.0030&0.0039&0.0054&0.0059&0.0080\\
     incompatibility ($\times10^{-5}$)&0.0074&0.0790&0.2864&0.6089&1.0610&2.0247&2.4242&4.3883\\
     \hline
     $P$ value ($10^{\wedge}$)&-1117&-3446&-6222&-8345&-8977&-10598&-80949&-20601\\
     \hline
\end{tabular}
\label{tab:result}
\end{table}

\subsection{Test of local realism}\label{lr}
To convincingly quantify the results of the experiment, we perform a hypothesis test of local realism. The null hypothesis is that the experimental results can be explained by local realism. We use the $p$ value obtained from test statistics to denote the maximum probability that the observed experimental results can be obtained from local realism. Hence, we have to show that our results correspond to a very small $p$ value so that they can indicate a strong rejection of local hidden-variable models. We apply the prediction-based radio (PBR) method to design the test statistics and compute the upper bound of the $p$ value. 

Denote Alice’s and Bob’s random settings distribution at each trial as $p_{A}(x)$ and $p_{B}(y)$ where $x,y\in\{0,1\}$. We assume the joint probability distribution of them, $p_{xy}$, is fixed and known before running the test. The measurement outcomes of Alice and Bob at each trial are denoted by A and B with possible value $a,b\in\{0,1,u\}$. In practice, we can only obtain the frequency of different outcomes $\textbf{f}\equiv\{n(abxy)/N,a,b=0,1,u;x,y=0,1\}$, where n is the number of trials the result $(a,b,x,y)$ appears and N is the number of the total experimental trials.
So, at the beginning of the test, we have to use the maximum likelihood method to find out the no-signaling distribution $\textbf{P}_{\text{NS}}^{*}\equiv\{p_{xy}p_{\text{NS}}^{*}(ab|xy),a,b=0,1,u;x,y=0,1\}$ that has the minimum distance from the observed frequency distribution $\textbf{f}$. When the number of trials is large enough, $\textbf{P}_{\text{NS}}^{*}$ can be close to the true probability distribution.

Particularly,  we use the Kullback-Leribler (KL) Ref.~\cite{kullback1951information} divergence to measure this distance because it’s the optimal 
asymptotic efficiency for rejecting the null hypothesis.
\begin{equation} \label{10}
    D_{\text{KL}}(\textbf{f}||\textbf{P}_{\text{NS}})=\sum_{a,b,x,y}p_{xy}f(ab|xy)\log_2\left(\frac{f(ab|xy)}{p_\text{NS}(ab|xy)}\right).
\end{equation}
Hence, the optimal no-signaling distribution $\textbf{P}_\text{NS}^*$ is the solution of the optimization
\begin{equation} \label{11}
    \min_{\textbf{P}_\text{NS} \in P_\text{NS}}D_\text{KL}(\textbf{f}||\textbf{P}_\text{NS}).
\end{equation}
After this estimation, we can find the optimal local realistic distribution $\textbf{P}_\text{LR}^*$ that has the minimum distance from our nearly true probability distribution.
\begin{equation} \label{12}
    \textbf{P}_\text{LR}^*=\mathop{\arg\min}\limits_{\textbf{P}_\text{LR} \in P_\text{LR}}D_\text{KL}(\textbf{P}^*_\text{NS}||\textbf{P}_\text{LR})=\sum_{a,b,x,y}p_{xy}p^*_\text{NS}(ab|xy)\log_2\left(\frac{p^*_\text{NS}(ab|xy)}{p_\text{LR}(ab|xy)}\right).
\end{equation}

Once the optimal no-signaling distribution $P_\text{NS}^*$ and local realistic distribution $P_\text{LR}^*$ are found according to the method in Ref.~\cite{Zhang2011Asymptotically}, any local realistic distribution $P_\text{LR}$ satisfies the following inequality:
\begin{equation}
    \sum_{a,b,x,y}\frac{p^*_\text{NS}(ab|xy)}{p_\text{LR}^*(ab|xy)}p_{xy}p_\text{LR}(ab|xy)\le1    
\end{equation}
The test statistics $R(ABXY)\equiv\frac{p^*_\text{NS}(ab|xy)}{p_\text{LR}^*(ab|xy)}$ provide a way to perform a hypothesis test of the local realism principle. To perform the hypothesis test without assuming the trial results are i.i.d., before the $i$-th trial we need to construct the test statistics $R_i(A_iB_iX_iY_i)$ for this trial. For this purpose, we need to replace the experimentally observed frequency distribution $\textbf{f}$ in Eqs.~\ref{10}, ~\ref{11} and the corresponding $P_\text{NS}^*$ in ~\ref{12} by a frequency distribution $\textbf{f}_i$ estimated before the $i$-th trial. The frequency distribution $\textbf{f}_i$ can be estimated using all the trial results before the $i$-th trial or using only the most recent trial results in history. Hence, the test statistics $R_i(A_iB_iX_iY_i)$ are called "prediction-based ratios", abbreviated as PBRs. Once the PBRs are constructed, after n trials the p-value upper bound for rejecting the no-signaling principle is given by:
\begin{equation}
    p_n=\min\left(\left(\prod_{i=1}^nR_i(A_iB_iX_iY_i)\right)^{-1},1\right)    
\end{equation}
where $a_ib_i$ and $x_iy_i$ are the measurement outcomes and setting choices at the $i$-th trial.
The corresponding $P$ values of the eight Bell violation are shown in Tab.~\ref{tab:result}

\subsection{interplay among nonlocality, entanglement and incompatibility}
\begin{figure}[htbp]
\centering
\includegraphics[width =0.7\textwidth]{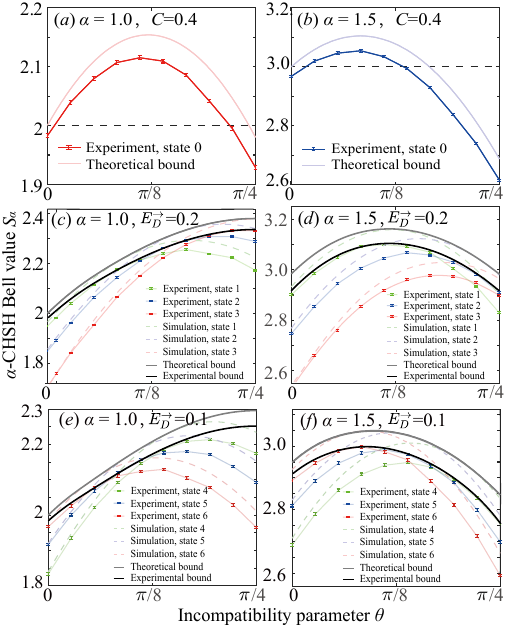}
\caption{The trajectory of the interplay between nonlocality and incompatibility. Subfigures (a)(b)(c)(d) have been discussed in detail in the main body of the text. In subfigures (e) and (f), the interplay trajectory is represented by a dark-colored line. This trajectory is derived from the convex hull from states 4, 5, and 6, depicted by faint line segments along with error bars. The upper bounds simulated for these states are presented as dashed lines, while the theoretical interplay trajectory is illustrated by a faint gray line.
}
\label{supple_interplay}
\end{figure}

 In our experimental setup, we select states with the concurrence $C=0.4$ and ODE $E_D^\rightarrow=0.1$, $E_D^\rightarrow=0.2$, and we experimentally evaluate the $\alpha$-CHSH value for $\alpha=1$ and $\alpha=1.5$ case when altering incompatibility, separately. The full results are shown in Fig.~\ref{supple_interplay}.

\subsection{Intersection correction}
\begin{figure}[htbp]
\centering
\includegraphics[width =0.5\textwidth]{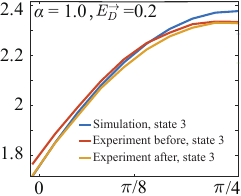}
\caption{the deviation between simulation trajectory and the experiment result before and after calibrating the optical axis of Pockels cell and HWP}
\label{supple_deviation}
\end{figure}

When $\alpha=1$, $E_D^\rightarrow=0.2$, upon observing the given figure, a slight translation on the curve is evident. Intuitively, At the beginning of the curve of state 3, there exists a point where the observed $\alpha$-CHSH Bell value surpasses the expected Bell value, which is the largest value achievable in this situation according to Eq.~\eqref{eq:interplay_optm_fixed_E}. From simulations, we observe that the presence of white noise in the state (resulting in a decrease in fidelity) only diminishes the measured value. Consequently, we suggest that there is a horizontal shift due to deviations in the measurement angles, which we attribute to variations in the optical axis of the polarization-changing devices. To validate this hypothesis, we meticulously calibrate the optical axis of the Pockels cell and HWP before repeating the experiment. The results of this verification are illustrated in Fig.~\ref{supple_deviation}.

Upon examination of the figure, a distinct horizontal translation of the curve is evident before and after the optical axis calibration. Following the calibration, the experimental curve aligns more closely with the theoretical curve.

Our experiments provide additional evidence supporting our simulation hypothesis that measurement deviations induce the horizontal shift, whereas, independently, state fidelity deviations, attributed solely to white noise, lead to the vertical shift. We believe that this methodology enables us to benchmark the measurement devices and state fidelity distinctly and with precision.

\end{appendix}
\bibliography{Interplay_arxiv}
\end{document}